\documentclass[letterpaper, 10 pt, conference]{ieeeconf}  

\IEEEoverridecommandlockouts
\usepackage{microtype}

\usepackage[scaled=.8]{beramono} 
\usepackage{amsmath} 
\usepackage{amssymb}  
\usepackage[ruled,vlined,linesnumbered]{TemplateFiles/algorithm2e}
\usepackage[usenames]{color}
\usepackage[svgnames]{xcolor}
\usepackage{mathrsfs}
\usepackage{graphicx}
\usepackage{amsfonts}
\usepackage{array}
\usepackage{flafter}
\usepackage{cite}
\usepackage{verbatim}
\usepackage{psfrag}
\usepackage{umoline}
\usepackage{epstopdf}
\usepackage{subcaption} 
\usepackage{hyperref}
\usepackage[capitalize]{cleveref}
\crefformat{equation}{(#2#1#3)}
\Crefformat{equation}{Equation~(#2#1#3)}
\Crefname{equation}{Equation}{Equations}
\usepackage{lipsum}

\hypersetup{%
    pdfborder = {0 0 0}
}
\graphicspath{{./figs/}}

\DeclareMathOperator*{\argmax}{arg\,\!max}

\newcommand{\boldblue}[1]{#1}

\crefname{algocf}{Alg.}{Algs.}
\Crefname{algocf}{Algorithm}{Algorithms}

\title{\LARGE \bf
Scalable Accelerated Decentralized Multi-Robot Policy Search \\in Continuous Observation Spaces
}

\author{Shayegan Omidshafiei$^{1}$, Christopher Amato$^{2}$, Miao Liu$^{3}$, Michael Everett$^{1}$, Jonathan P. How$^{1}$, John Vian$^{4}$
	\thanks{*This work was supported by The Boeing Company.}
	\thanks{$^{1}$Laboratory for Information and Decision Systems (LIDS), MIT, Cambridge, MA 02139, USA {\tt\small \{shayegan,jhow\}@mit.edu}}%
	\thanks{$^{2}$College of Computer and Information Science (CCIS), Northeastern University, Boston, MA 02115, USA {\tt\small camato@ccs.neu.edu}}%
	\thanks{$^{3}$IBM Thomas J. Watson Research Center, Yorktown Heights, NY 
		10598, USA {\tt\small miao.liu1@ibm.com }(work completed while the author was at MIT)}%
	\thanks{$^{4}$Boeing Research \& Technology, Seattle, WA 98108, USA {\tt\small john.vian@boeing.com}}%
}

\newcounter{Lcount}

\renewcommand{\baselinestretch}{0.97}

\begin{document}
\maketitle
\thispagestyle{empty}
\pagestyle{empty}

\begin{abstract}
This paper presents the first ever approach for solving \emph{continuous-observation} Decentralized Partially Observable Markov Decision Processes (Dec-POMDPs) and their semi-Markovian counterparts, Dec-POSMDPs. This contribution is especially important in robotics, where a vast number of sensors provide continuous observation data. A continuous-observation policy representation is introduced using Stochastic Kernel-based Finite State Automata (SK-FSAs). An SK-FSA search algorithm titled Entropy-based Policy Search using Continuous Kernel Observations (EPSCKO) is introduced and applied to the first ever continuous-observation Dec-POMDP/Dec-POSMDP domain, where it significantly outperforms state-of-the-art discrete approaches. This methodology is equally applicable to Dec-POMDPs and Dec-POSMDPs, though the empirical analysis presented focuses on Dec-POSMDPs due to their higher scalability. To improve convergence, an entropy injection policy search acceleration approach for both continuous and discrete observation cases is also developed and shown to improve convergence rates without degrading policy quality. 


\end{abstract}
\section{Introduction} \label{sec:intro}

Decision-making under uncertainty is a ubiquitous robotics problem wherein a robot collects data from its environment and decides subsequent tasks to execute. While low-cost robotics platforms and sensors have increased the affordability of multi-robot systems, derivation of policies dictating robot decisions remains a challenge. This decision-making problem is even more complex in noisy settings with imperfect communication, requiring a formal framework for its treatment.

A general representation of the multi-agent planning under uncertainty problem is the Decentralized Partially Observable Markov Decision Process (Dec-POMDP) \cite{Bernstein02}, which extends single-agent POMDPs to decentralized domains. Due to Dec-POMDPs' usage of primitive actions (atomic actions assumed to each take a single time unit to execute) they have exceedingly large policy spaces which severely limits planning scalability. Recent efforts have extended Dec-POMDPs to use macro-actions (temporally extended actions), resulting in the Decentralized Partially Observable \emph{Semi}-Markov Decision Process (Dec-POSMDP) \cite{AmatoRSS15_v2,Omidshafiei15_ICRA}. The result is a scalable asynchronous multi-robot decision-making framework which plans over the space of high-level robot tasks (e.g., \emph{Open-the-valve} or \emph{Find-the-key}) with non-deterministic durations. 

Despite the increased action-space scalability offered by Dec-POSMDPs, they have so far been limited to planning over the space of discrete observations. To date, no algorithms exist for continuous-observation Dec-POSMDPs (or Dec-POMDPs \cite{DecPOMDPBook16}). This is a major research gap, especially important in the context of robotics where a vast number of real-world sensors provide continuous observation data. 
Application of Dec-POSMDPs to continuous problems such as robot navigation currently mandates observation space discretization, resulting in loss of valuable sensor information which could otherwise be used to better inform the decision-making policy. Several approaches have targeted single-agent continuous-observation POMDPs. These include partitioning of continuous spaces into lossless discrete spaces \cite{hoey2005solving}, Gaussian mixtures for belief representation \cite{porta2006point}, use of continuous-observation classifiers \cite{bai2014integrated}, and learned discrete representations for continuous state spaces \cite{brechtel2013solving}. This paper expands this body of work beyond the single-agent case, targeting scalable treatment of continuous-observation Dec-POSMDPs. The methods presented are applicable to domains with continuous underlying state spaces, as shown in some of the experiments used for evaluation.

In order to develop solvers for continuous-observation Dec-POSMDPs, we build on current state-of-the-art discrete policy search methods  \cite{Omidshafiei16_ICRA,AmatoRSS15_v2,Omidshafiei15_ICRA}. Unfortunately, these algorithms suffer from convergence speed limitations---an issue which was identified in prior work but remains untreated \cite{Omidshafiei16_ICRA}. A major gap exists in addressing these issues before extending the foundations of these discrete algorithms to the continuous case, where such convergence issues are exacerbated. To resolve this, we first introduce a maximal entropy injection approach targeting convergence acceleration for both discrete and continuous algorithms, without degrading overall policy quality. The approach is shown to significantly outperform existing search acceleration methods.

The paper's key contribution is a stochastic kernel-based policy representation and search algorithm, allowing direct mapping of continuous observations to robot decisions (with no discretization necessary). This algorithm leverages the proposed entropy injection acceleration method and is evaluated on a multi-robot nuclear contamination domain---the first ever continuous-observation Dec-POMDP/Dec-POSMDP domain---in which discrete policy search algorithms perform extremely poorly. Failure modes of discrete methods are analyzed and compared to the superior continuous policy behavior. The contributions introduced in this paper can be readily applied to Dec-POMDPs \emph{and} Dec-POSMDPs. However, as we are motivated by applications to extremely large action-observation spaces, the notation used and experiments conducted focus on the more scalable Dec-POSMDP framework.

\section{Background} \label{sec:background}
\subsection{Decentralized Planning using Macro-Actions}\label{subsec:decposmdp}
This section summarizes the Dec-POSMDP, a multi-robot decentralized decision-making under uncertainty framework targeting action-space scalability. For a more detailed introduction to Dec-POSMDPs, we refer readers to \cite{Omidshafiei16_ICRA,AmatoRSS15_v2,Omidshafiei15_ICRA}.

The Dec-POSMDP is a belief-space framework in which agents execute macro-actions (temporally-extended actions) with non-deterministic completion times, and receive noisy high-level observations of their post-MA state. Macro-actions (MAs) are abstractions of low-level POMDPs involving primitive actions $u_{t}^{(i)}$ and observations $o_{t}^{(i)}$, allowing execution of high-level tasks (e.g., \emph{Park-the-car})\footnote{We denote a generic parameter $p$ of the $i$-th robot as $p^{(i)}$, a joint team parameter as $\bar{p}$, and a joint team parameter at timestep $k$ as $\bar{p}_{k}$.}. Each MA executes until an $ \epsilon $-neighborhood of its belief milestone $\check{b}^{goal}$ is reached. This neighborhood defines the MA termination condition or \emph{goal belief node}, denoted $ B^{goal}\!=\!\{b~:~\|b-\check{b}^{goal}\|\leq\epsilon \} $ \cite{Omidshafiei15_ICRA}.

Upon completion of an MA, each robot makes a macro (or high-level) observation $o^{e(i)}$ of the underlying high-level system state $x^{e} \in \mathbb{X}^{e}$. It also calculates its own final belief state, $b^{f(i)}$. Thus far, both Dec-POMDPs and Dec-POSMDPs have only seen limited applications to \emph{finite discrete} observation spaces. Due to its action-space scalability, let us focus on the Dec-POSMDP, defined as follows:
\begin{itemize}
	\item $\mathbb{I}=\{1,2,\ldots,n\}$ is the set of heterogeneous robots.
	\item $\mathbb{B}^{(1)}\times\mathbb{B}^{(2)}\times\ldots\times\mathbb{B}^{(n)}\times\mathbb{X}^{e} $ is the belief space, with local belief milestones $ \mathbb{B}^{(i)} $ and joint environment (or high-level) space $\mathbb{X}^{e}$.
	\item $ \bar{\mathbb{T}}=\mathbb{T}^{(1)}\times\mathbb{T}^{(2)}\ldots\times\mathbb{T}^{(n)} $ is the joint MA space, where $ \mathbb{T}^{(i)} $ is the finite set of MAs for the $ i $-th robot. 
	\item $ \bar{\breve{\mathbb{O}}}^{e}=\{\bar{\breve{o}}^{e} \} $ is the space of all joint MA-observations. 
	\item $ P(\bar{b}',x^{e'},k|\bar{b},x^{e};\bar{\pi}) $ is the high-level transition probability model under MAs $ \bar{\pi} $ from $ (\bar{b},x^{e}) $ to $ (\bar{b}',x^{e'}) $.
	\item $ \bar{R}^{\tau}\!(\bar{b},x^{e};\bar{\pi}) $ is the high-level reward of taking a joint MA $ \bar{\pi} $ at $(\bar{b},x^{e})$. 
	\item $ P(\bar{\breve{o}}^{e}|\bar{b},x^{e}) $ is the joint observation likelihood model, with joint observation $\bar{\breve{o}}^{e}=\{\breve{o}^{e(1)},\breve{o}^{e(2)},\ldots,\breve{o}^{e(n)} \} $.
	\item $\gamma \in [0, 1)$ is the reward discount factor.
\end{itemize}

Macro-observations and final beliefs are jointly denoted as \emph{MA-observation} $\breve{o}^{e(i)} = (o^{e(i)}, b^{f(i)})$. Trajectories of MAs and received MA-observations are denoted as the \emph{MA-history},
\begin{align}
\xi^{(i)}_{k}=\{\breve{o}^{e(i)}_{0},\pi^{(i)}_{0},\breve{o}^{e(i)}_{1},\pi^{(i)}_{1},\ldots,\breve{o}^{e(i)}_{k-1},\pi^{(i)}_{k-1},\breve{o}^{e(i)}_{k}\}.
\end{align}
Transition probability $P(\bar{b}'\!\!,x^{e'}\!\!\!,k|\bar{b},x^{e};\!\bar{\pi})$ from $(\bar{b},x^{e}\!)$ to $(\bar{b}'\!,x^{e'}\!)$ under joint MA $\bar{\pi} \!=\! \{\pi^{(1)}\!,\!\ldots\!,\pi^{(n)}\}$ in $k$ timesteps is,
\vspace{-5pt}
\begin{align} \label{eq:MA_trans_prob_epoch}
& P(\bar{b}',x^{e'},k|\bar{b}_{0},x^{e}_{0},o^{e}_{k};\bar{\pi})
= P(x^{e}_k,\bar{b}_k|\bar{b}_{0},x^{e}_{0},o^{e}_{k};\bar{\pi})\nonumber\\
&~~~~~~=\sum_{x^{e}_{k-1},\bar{b}_{k-1}}\Big[P(x^{e}_k|x^{e}_{k-1},o^{e}_{k};\bar{\pi}(\bar{b}_{k-1}))\times \\
&P(\bar{b}_k|x^{e}_{k-1},\bar{b}_{k-1};\bar{\pi}(\bar{b}_{k-1})) P(x^{e}_{k-1},\bar{b}_{k-1}|x^{e}_{0},\bar{b}_{0};\bar{\pi}(\bar{b}_{0}))\Big].\nonumber
\end{align}

The generalized high-level team reward for a discrete-time Dec-POSMDP during execution of joint MA $\bar{\pi}$ is defined \cite{Omidshafiei16_ICRA},
\begin{align}\label{eq:gen_reward_k=0}
\!\!\!\!\!\bar{R}^{\tau}\!(\bar{b},x^{e};\bar{\pi}) \!=\! \mathbb{E}\!\left[\!\sum_{t=0}^{\tau-1}\!\!\gamma^{t}\!\bar{R}(\bar{x}_{t},x^{e}_{t}\!,\bar{u}_{t})|P(\bar{x}_{0})\!=\!\bar{b},x^{e}_{0}\!=\!x^{e}\!;\bar{\pi}\!\right]
\end{align}
where $\tau = \min_{i}\min_{t}\{t:b^{(i)}_{t}\in B^{(i),goal} \}$ is the first timestep at which any robot completes its current MA.

The \emph{joint high-level policy}, $\bar{\phi} = \{\phi^{(1)}, \ldots, \phi^{(n)} \}$, dictates MA selection. High-level policy $\phi^{(i)}$ maps the $i$-th robot's MA-history $\xi^{(i)}_{k}$ to the next MA $\pi^{(i)}$ to be executed. Joint Dec-POSMDP value under policy $\bar{\phi}$ is then \cite{Omidshafiei16_ICRA},
\begin{align}\label{eq:decposmdp_eval}
\!\bar{V}^{\bar{\phi}}(\bar{b},x^{e})&=\mathbb{E}\left[\sum_{k=0}^{\infty}\gamma^{t_{k}}\bar{R}^{\tau}(\bar{b}_{t_{k}},x^{e}_{t_{k}};\bar{\pi}_{t_{k}})|\bar{b}_{0},x^{e}_{0};\bar{\phi}\right]\\
&=\bar{R}^{\tau}\!(\bar{b},x^{e};\bar{\pi}) \notag + \\
& \sum_{k=1}^\infty\gamma^{t_k}\!\!\!\!\!\!\sum_{\bar{b}',x^{e'},o^{e'}}\!\!\!\!\!\!P(\bar{b}'\!,x^{e'}\!\!,o^{e'}\!\!,k|\bar{b},x^{e};\bar{\pi})\bar{V}^{\bar{\phi}}(\bar{b}'\!,x^{e'}).
\end{align}
The optimal joint high-level policy is,
\begin{align}\label{eq:decposmdp_problem}
\bar{\phi}^{*}=\argmax_{\bar{\phi}}\bar{V}^{\bar{\phi}}(\bar{b},x^{e}).
\end{align}

Solving the Dec-POSMDP results in joint high-level decision-making policy $\bar{\phi}$ dictating the MA $\pi^{(i)}$ executed by each robot based on its MA-history. Each MA is, itself, a policy over low-level actions $u_t^{(i)}$ and observations $o_t^{(i)}$. Thus, decision-making using the Dec-POSMDP allows abstraction of task-level actions from low-level actions, leading to significantly improved planning scalability over Dec-POMDPs. 

\subsection{Dec-POSMDP Policy Search Algorithms}\label{subsec:decposmdp_solns}
So far, research efforts have focused on Dec-POSMDP policy search for discrete observation spaces, resulting in several algorithms: Masked Monte Carlo Search (MMCS) \cite{Omidshafiei15_ICRA}, MacDec-POMDP Heuristic Search (MDHS) \cite{AmatoRSS15_v2}, and Graph-based Direct Cross Entropy method (G-DICE) \cite{Omidshafiei16_ICRA}. These algorithms use Finite State Automata (FSAs) for policy representation. FSA-based policy $\phi^{(i)}$ for robot $i$ consists of $N_n$ FSA nodes, $\{q^{(i)}_1,\ldots,q^{(i)}_{N_n}\}$. FSA-based decision-making is two-fold: each robot begins execution in FSA node $q^{(i)}$, where \emph{MA output function} $\pi^{(i)} = \lambda^{(i)}(q^{(i)})$ assigns it an MA, $\pi^{(i)}$. Following MA execution, the robot receives a high-level observation and selects its next FSA node using \emph{transition function} $q'^{(i)} = \delta^{(i)}(q^{(i)},\breve{o}^{e(i)})$. The graph-based nature of FSAs allows their application to infinite-horizon domains.

Though Dec-POSMDPs have increased the size of solvable planning domains beyond Dec-POMDP counterparts, major algorithm limitations still exist. MMCS is a greedy algorithm which succumbs to local optimality issues \cite{Omidshafiei15_ICRA}. MDHS uses lower and upper bound value heuristics to bias search towards promising policy regions, by initiating an empty (partial) FSA and incrementally assigning nodes actions $\lambda^{(i)}$ and transitions $\delta^{(i)}$. Partial policies with high upper bounds are expanded incrementally. Yet, each expansion involves $|\mathbb{T}| N_n^{|\mathbb{\bar{\breve{O}}}|}$ child policies, severely limiting usage for large observation spaces. 

G-DICE is a cross entropy-based algorithm which iteratively updates policies using two \emph{sampling distributions} at each FSA node: MA distribution $f(\pi^{(i)}|q^{(i)};\theta^{(i)(\pi|q)})$ and node transition distribution $f(q^{(i)'}|q^{(i)},\breve{o}^{e(i)};\theta^{(i)(q'|q,\breve{o}^e)})$, where $\theta^{(i)}$ are parameter vectors. Each iteration samples the distributions $N_s$ times, resulting in $N_s$ deterministic FSA policies. Maximum likelihood estimates (MLE) of parameters $\theta^{(i)}$ are calculated using the $N_b\!\leq\! N_s$ best policies. To prevent convergence to local optima, smooth parameter updates,
\begin{equation}\label{eq:gdice_smoothing}
\theta_{k+1} \gets \alpha\theta_{k+1} + (1-\alpha)\theta_{k},
\end{equation}
are used, with iteration number $k$ and learning rate $\alpha \in (0,1]$. For sufficiently small values of $\alpha$, this process minimizes cross entropy between each sampling distribution and a unit mass centered at the optimal policy \cite{journals/orl/CostaJK07}. G-DICE is executed until convergence, after which the best \emph{deterministic} policy from the history of samples is returned.

Using smooth parameter updates and sampling distributions initiating from a uniform distribution allows G-DICE to tradeoff exploration and exploitation in the policy space, outperforming other Dec-POSMDP search approaches given a fixed computational budget. Yet, G-DICE suffers from sample degeneracy and convergence issues related to the sampling distributions, and in its current form only applies to discrete observation settings. The following sections resolve these issues, resulting in a scalable, accelerated continuous-observation search algorithm.

\section{Accelerated Policy Search}\label{sec:max_ent_injection}
Prior to extending to continuous observations, this section treats the sampling distribution degeneracy issue in sampling-based Dec-POSMDP approaches. It also introduces a maximal entropy injection scheme which is then embedded in the proposed continuous-observation Dec-POSMDP algorithm. 

\subsection{Sampling Distribution Degeneracy Problem}\label{subsec:prev_approaches}
A major issue with sampling distribution-based approaches, such as G-DICE, occurs when a low enough learning rate $\alpha$ is not used, causing underlying sampling distributions to rapidly converge to \emph{degenerate distributions} far from the optimum \cite{conf/wsc/BotevK04}. All subsequent search iterations return identical samples of the policy space, stifling exploration altogether. Yet, one benefit of a high learning rate is fast convergence, especially useful for complex Dec-POSMDPs with large observation spaces and computationally expensive trajectory sampling and evaluation. Sampling distribution-based approaches such as G-DICE often require \emph{hand-tuned} selection of $\alpha$ for good performance, even after which convergence may be excessively slow and can hinder experimentation and analysis. This trade-off was noted in \cite{Omidshafiei16_ICRA}, where it was left as future work. Recall the motivation behind the Dec-POSMDP framework is scalability to very large multi-robot planning domains. Despite the fact that policy search is conducted offline, hindrance of human-in-the-loop analysis due to slow convergence is undesirable. A na\"{i}ve solution is to set $\alpha$ arbitrarily low, but this implies arbitrarily high convergence time (on the order of many days for complex domains). These foundational issues must first be resolved before extending these algorithms to treat the more complex continuous observation case.

Several works have targeted this degeneracy problem. One approach uses dynamic smoothing of learning rates \cite{Kroese2006}, 
\begin{equation}
\alpha_k = \alpha_0 - \alpha_0 (1-k^{-1})^{\beta},
\end{equation}
where $\alpha_0$ is the baseline rate (typically close to $1$) and $\beta$ is the drop-off rate (typically between $5$ to $10$). The result is a monotonically decreasing $\alpha_k$ which initially starts high.  

Another approach involves the addition of a noise term $\omega_k$ to the sampling distribution at each iteration $k$ to prevent degeneration. Linearly decreasing noise injection,
\begin{equation}
\omega_k = \max(\omega_{\rm max}-rk,0),
\end{equation}
was investigated in \cite{journals/icga/ThieryS09a}. In the above,  $\omega_{\rm max}$ is the maximum allowable noise and $r$ is the noise drop-off rate. 

These approaches are not ideal as they are agnostic to Dec-POSMDP value function convergence, meaning they do not adapt to domain-specific behaviors. Thus, sub-parameters ($\alpha_0$, $\beta$, $\omega_{\rm max}$, $r$) typically need significant tuning to alleviate convergence issues for individual domains. 

\subsection{Maximal Entropy Injection}\label{subsec:max_entropy_injection}

\begin{figure*}[t!]
	\centering
	\includegraphics[trim={0 0.2cm 0 0cm},width=1\linewidth]{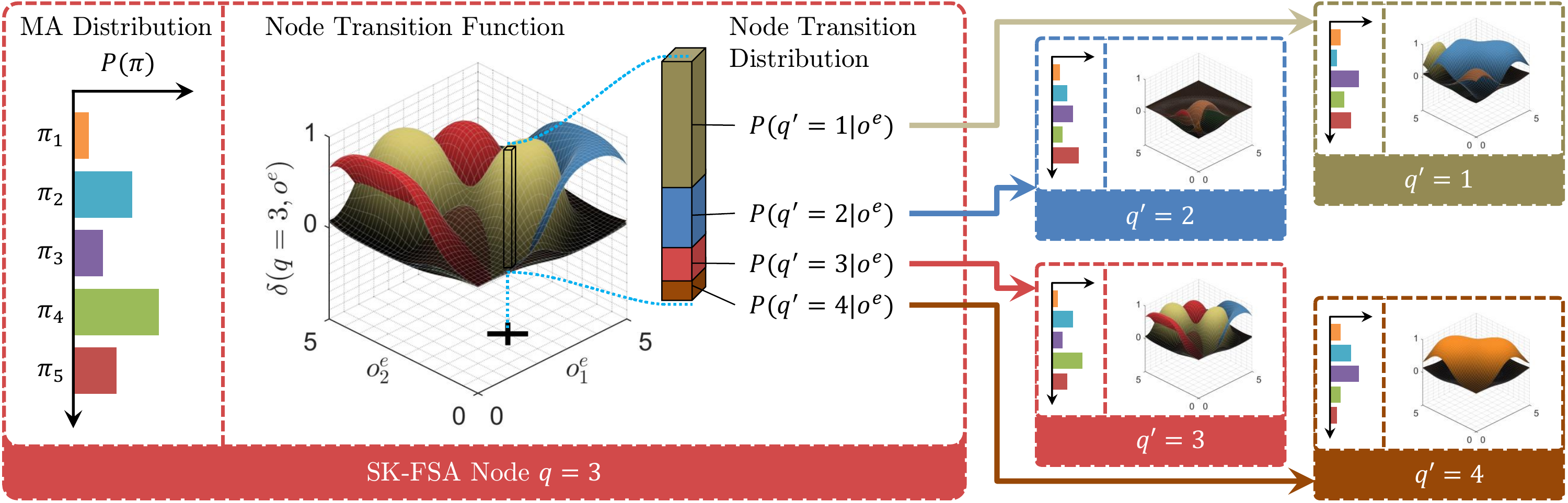}
	\begin{minipage}[t]{.12\linewidth}
		\centering
		\vspace{-4pt}
		\subcaption{The robot samples an MA using its node's MA distribution.}\label{fig:sk_fsa_overview_1}
	\end{minipage}%
	\hspace*{\fill}
	\begin{minipage}[t]{.47\linewidth}
		\centering
		\vspace{-4pt}
		\subcaption{Given a high-level continuous observation made following MA execution (e.g., $o^e = (2.5,1.5)$ above), the transition \emph{function} outputs a categorical transition \emph{distribution} over next-nodes $q'$. The robot samples this distribution to select its next node, $q'$.}\label{fig:sk_fsa_overview_2}
	\end{minipage}
	\hspace*{\fill}
	\begin{minipage}[t]{.36\linewidth}
		\centering
		\vspace{-4pt}		
		\subcaption{The robot repeats this decision-making process at the next SK-FSA node $q'$ (one of the 4 nodes above), conducting the stochastic MA and transition selection process indefinitely.}\label{fig:sk_fsa_overview_3}
	\end{minipage}
	\vspace{-8pt}
	\caption{Overview of continuous-observation decision-making using SK-FSAs. A given robot's policy is represented by a set of stochastic FSA nodes, each containing an MA sampling distribution and node transition function. A 4 node SK-FSA ($N_n = 4$) is illustrated above, with the robot starting policy execution at SK-FSA node $q=3$ (on the left).} \label{fig:sk_fsa_overview}
	\vspace{-15pt}
\end{figure*}

A principled approach combining policy exploration with fast convergence is desired, without reliance on sensitive dynamic smoothing or noise terms. As degenerate distributions have minimal entropy \cite{devroye2013probabilistic}, an intuitive idea is to simultaneously monitor policy value convergence and underlying sampling distribution entropy to alleviate degeneracy issues.

In the proposed acceleration approach, search is conducted as usual for iterations where policy value has not converged, allowing policy space exploration. Once convergence occurs, entropies of sampling distributions $f(\pi^{(i)}|q^{(i)};\theta^{(i)(\pi|q)})$ and $f(q^{(i)'}|q^{(i)},\breve{o}^{e(i)};\theta^{(i)(q'|q,\breve{o}^e)})$ are calculated. If a distribution's entropy is significantly below the max entropy for its distribution family, degeneracy has likely occurred \cite{devroye2013probabilistic}. Max entropy distributions are well-studied and closed form results for many families and constraint sets are known \cite{shannon2001mathematical}. For Dec-POSMDPs, these entropy calculations are computationally cheap as sampling distributions are categorical, with corresponding discrete uniform maximal entropy distributions. 

In post-degeneracy iterations, each sampling distribution's entropy is increased by incrementally combining its parameters $\theta^{(i)}$ with the max entropy distribution parameters $\theta_{ME}$, 
\begin{equation}\label{eq:entropy_injection}
\theta_{k+1} \gets (1-\alpha_{EI})[\alpha\theta_{k+1} + (1-\alpha)\theta_{k}] + \alpha_{EI}\theta_{ME},
\end{equation}
where $\alpha_{EI}$ is the entropy injection rate. This encourages policy space exploration while still allowing usage of high learning rates (e.g., $\alpha > 0.5$) for fast convergence. In practice, entropy injection rate $\alpha_{EI}$ has a low value (between 1\% - 3\% per iteration). As this process is repeated \emph{only} in post-convergence iterations, there is low sensitivity to $\alpha_{EI}$ as entropy is incrementally increased \emph{whenever necessary}. Injection stops as soon as the policy value diverges, allowing unhindered exploration. This acceleration approach is evaluated in \cref{subsec:expts_accel} and also integrated into the proposed continuous-observation search algorithm in the next section.

\section{Continuous-Observation Dec-POSMDP Search}\label{sec:continuous_obs}
This section focuses on multi-robot policy search in continuous observation spaces. It first presents an extension of traditional discrete, deterministic FSAs to allow representation of continuous policies. A continuous-observation Dec-POSMDP search algorithm is then introduced.

\subsection{Stochastic Kernel-Based Finite State Automata}\label{subsec:sk-fsas}
We first extend the notion of deterministic policies used in existing Dec-POSMDP algorithms to stochastic policies. In a stochastic FSA, MA output function $\lambda^{(i)}$ and node transition function $\delta^{(i)}$ provide robots with a probability distribution over MAs and next-nodes $q'$ during policy execution, rather than deterministic MA and transition assignments. The resulting stochastic decision-making scheme allows robots to escape cycles of incorrect decisions which may otherwise occur in deterministic FSAs \cite{amato2010optimizing}. 
While it has been shown that finite-horizon Dec-POMDPs have at least one optimal deterministic policy (i.e., guaranteed to at least equal performance of the optimal stochastic policy) \cite{oliehoek2010value}, in approximate searches, stochastic FSAs often result in a higher joint value \cite{bernstein2009policy, amato2010optimizing}. One can readily modify cross entropy-based search to provide such a stochastic policy by simply using the underlying sampling distributions $f(\pi^{(i)}|q^{(i)};\theta^{(i)(\pi|q)})$ and $f(q^{(i)'}|q^{(i)},\breve{o}^{e(i)};\theta^{(i)(q'|q,\breve{o}^e)})$ to define the policy, rather than the \emph{best sampled deterministic policy} (as done in G-DICE).


A second issue is extension of FSAs to support continuous observations, a formidable task as continuous observation spaces are uncountably infinite. Existing Dec-POSMDP algorithms are, thus, inapplicable. To resolve this, we assume policy smoothness over the observation space, a characteristic which occurs naturally in many robotics domains. In other words, the controller structure should induce similar decisions from similar observation chains. This typical assumption is also made by the continuous state-action MDP and POMDP literature \cite{carden2014convergence,brechtel2013solving,bai2014integrated}. 

We exploit this smoothness assumption and introduce Stochastic Kernel-based Finite State Automata (SK-FSAs) for policy representation (\cref{fig:sk_fsa_overview}), which have similar structure to the controllers used in \cite{bai2014integrated}. Policy execution in SK-FSAs is similar to traditional FSAs. Each robot's SK-FSA node (e.g., node $q = 3$ in \cref{fig:sk_fsa_overview}) outputs categorical MA distribution $f(\pi^{(i)}|q^{(i)};\theta^{(i)(\pi|q)})$, which the robot samples to select its next MA (\cref{fig:sk_fsa_overview_1}). Following MA execution, the robot receives a \emph{continuous} high-level observation, which the SK-FSA node transition \emph{function} $\delta^{(i)}$ uses to output a corresponding node transition \emph{distribution} $f(q^{(i)'}|q^{(i)},\breve{o}^{e(i)};\theta^{(i)(q'|q,\breve{o}^e)})$. Note the distinction between transition \emph{function} and transition \emph{distribution}---the transition function maps continuous observations to the $N_n$-dimensional simplex. Given an observation, $\delta^{(i)}$ outputs an infinitesimal `slice' representing a categorical transition distribution over next-nodes $q'$. \cref{fig:sk_fsa_overview_2} illustrates such a slice, evaluated at high-level observation $o^e = (2.5,1.5)$. The robot samples this categorical distribution, transitions to its next SK-FSA node $q'$, and repeats this process indefinitely.

We propose use of kernel logistic regression (KLR) to represent node transition functions. KLR is a non-parametric multi-class classification model (i.e., model complexity grows with the number of kernel points). In SK-FSAs, node transition functions use KLR with high-level observation inputs, $o^e$, and output probabilities over next-nodes $q'$. KLR is a natural model for stochastic policies as it is a probabilistic classifier (i.e., SK-FSA transition distributions correspond to KLR probabilities) \cite{zhu2012kernel}. Our approach uses KLR with radial basis function (RBF) kernels over the observation space,
\begin{equation}
K(o^e,o^{e'}) = \exp(-0.5\sigma^{-2}||o^e - o^{e'}||^{2}),
\end{equation}
where $\sigma$ is the kernel radius. RBF kernels are preferred as they provide smooth classification outputs while allowing non-linear decision boundaries \cite{zhu2012kernel}, in contrast to linear kernels. The next section discusses SK-FSA policy search, including details on kernel basis selection and kernel weight training.

\subsection{Entropy-based Policy Search over SK-FSAs}\label{subsec:epscko}
This section introduces an SK-FSA search algorithm titled Entropy-based Policy Search using Continuous Kernel Observations (EPSCKO). EPSCKO consists of 3 steps: cross entropy search for MA distributions (as done in G-DICE), memory-bounded KLR training for SK-FSA node transition functions, and entropy injection for search acceleration (as in \cref{subsec:max_entropy_injection}). In each EPSCKO iteration, decision trajectories are sampled from the SK-FSA policy. The $N_b$ best trajectories (evaluated using \cref{eq:decposmdp_eval}) are used for policy update.

We first detail the KLR training approach and then present the overall algorithm. As transition function $\delta^{(i)}$ uses a kernel-based representation over the observation space, it requires a set of observation kernel basis points and weights. In EPSCKO, kernel weights constitute the node transition parameter vector $\theta^{(i)(q'|q,\breve{o}^e)}$. To simplify notation, references to $\theta_{q'}^{(i)}$ in this section refer to this transition parameter vector.

The computational cost of training KLR models is $O(N_d^3)$ \cite{zhu2012kernel}, where $N_d$ is the training input size. For a sustainable training time, EPSCKO uses a memory-bounded kernel basis consisting of continuous observations received during evaluation of the $N_b$ best policies in each of the latest $N_{KLR}$ iterations. In each iteration, the \emph{bundle} of observations in the $N_b$ best decision trajectories is pushed to a first-in, first-out (FIFO) circular queue of length $N_{KLR}$. KLR training outputs are the corresponding sampled node transitions taken along these same trajectories. The non-parametric nature of KLR ensures that node transition function complexity increases in regions with high observation density, so the policy naturally focuses on prominent observation space regions. The result is a compact yet informative policy representation.


\SetKwFunction{EPSCKO}{EPSCKO}
\begin{algorithm}[t!]
	\caption{EPSCKO}\label{alg:EPSCKO}
	\textbf{Procedure}: $ \bar{\phi}_{b}  = \EPSCKO(\bar{\mathbb{T}}, \bar{\breve{\mathbb{O}}}^{e}, \mathbb{I}, N_{n}, N_{k}, N_{s}, N_{b}, N_{KLR}, \alpha, \alpha_{EI})$\\
	{
		For each robot, initialize SK-FSA policy with $N_n$ nodes;\label{line:epscko_init_graph}\\
		
		\boldblue{$\bar{Q}_{KLR} \gets \text{initFIFOQueue}(N_{KLR})$;\label{line:epscko_init_queue}}\\
		$\bar{V}_{b}, \bar{V}_{w,0} \gets -\infty$;\label{line:epscko_init_V}\\
		
		\For {$i = 1$ to $n$} 
		{
			Initialize $\theta_{0}^{(i)(\pi|q)}~~\forall q$ and $\theta_{0}^{(i)(q'|q,\breve{o})}~~\forall q,\breve{o}$;\label{line:init_params_1}\\
		}
		
		\For{$k=0$ to $N_k-1$}
		{
			allowEntropyInject, entropyInjected $\gets$ \texttt{False};\label{line:allow_ent_inject}\\
			
			$\bar{\pi}_{list}, \overline{KLR}_{list}, \bar{V}_{list} \gets \varnothing$;\\
			\For{$s=1$ to $N_s$}
			{
				\boldblue{$\bar{V}^{\bar{\phi}}, \{\bar{\pi}\}_s, \{\bar{\breve{o}}^e, \bar{q}'\}_s \gets$ Evaluate$(\bar{\phi})$\label{line:decposmdp_eval};}\\
				
				\If {$\bar{V}^{\bar{\phi}} \geq \bar{V}_{w,k}$}{
					$\bar{\pi}_{list} \gets \bar{\pi}_{list} \cup \{\bar{\pi}\}_s$;\label{line:epscko_reject_bad_policies_1}\\
					$\overline{KLR}_{list} \gets \overline{KLR}_{list} \cup \{\bar{\breve{o}}^e, \bar{q}'\}_s$;\label{line:epscko_reject_bad_policies_2}\\
					$\bar{V}_{list} \gets \bar{V}_{list} \cup \bar{V}^{\bar{\phi}}$;\label{line:epscko_reject_bad_policies_3}\\
				}
				
				\If{$\bar{V}^{\bar{\phi}} > \bar{V}_b$}{
					$\bar{V}_{b}, \bar{\phi}_{b} \gets \bar{V}^{\bar{\phi}}, \bar{\phi}$;\label{line:epscko_save_best_policy}\\				
				}
			}
			\boldblue{$\bar{\pi}_{b,list}, \overline{KLR}_{b,list}, \bar{V}_{b,list} \!\gets$ best $N_b$ policies in $\bar{V}_{list}$};\label{line:epscko_retain_best_nb}\\
			\boldblue{$\bar{Q}_{KLR}.\text{push}(\overline{KLR}_{b,list})$};\label{line:epscko_append_oe_hist}\\
			$\bar{V}_{w,k+1} \gets$ min$(\bar{V}_{b,list})$;\\
			
			\boldblue{
				\If{ValueConverged()}{
					allowEntropyInject $\gets$ True;\label{line:epscko_allow_injection}
				}
			}
			
			\For {$i = 1$ to $n$}
			{
				$\theta_{k+1}^{(i)(\pi|q)} \gets$ MLE of $\theta^{(i)(\pi|q)}$ using $\bar{\pi}_{b,list}~~\forall q $;\label{line:epscko_mle_update}\\
				
				$\theta_{k+1}^{(i)(\pi|q)}\gets\alpha\theta_{k+1}^{(i)(\pi|q)}+(1-\alpha)\theta_{k}^{(i)(\pi|q)}$;\label{line:epscko_param_smooth}\\
				
				\boldblue{$\theta_{k+1}^{(i)(q'|q,\breve{o})} \!\! \gets \text{trainWeightedKLR}(Q^{(i)}_{KLR},\alpha)$;}\label{line:epscko_klr_update}\\
				
				\boldblue{
					\If{allowEntropyInject}{
						$\!$entropyInjected $\!\!\gets\!\!$ tryInject($\theta_{k+1}^{(i)(\pi|q)}\!\!\!, \theta_{k+1}^{(i)(q'|q,\breve{o})}$)\label{line:epscko_try_inject};
					}
				}
			}
			
			\boldblue{
				\If{entropyInjected}{
					$\bar{V}_{w,k+1} \gets$ $-\infty$;\label{line:epscko_reset_worst};
				}
			}
		}
		\Return $\bar{\phi}_{b}$;\\
	}
\end{algorithm}


To counter convergence to locally optimal SK-FSAs, EPSKCO uses a weighted log-likelihood function to train the KLR model. Weights are discounted such that observations sampled in earlier algorithm iterations are given higher value. Given learning rate $\alpha$, the following weight set is used,
\begin{equation}\label{eq:klr_mixture_weights}
w_{b} = \begin{cases}
(1-\alpha)^{N_{KLR}-1} & b=1\\
\alpha(1-\alpha)^{N_{KLR}-b} & b \in \{2,\cdots,N_{KLR}\}
\end{cases}
\end{equation}
where $w_{b}$ is the training weight for the $b$-th observation bundle in the FIFO kernel queue. This weighting is derived from recursive application of \eqref{eq:gdice_smoothing}, and is analogous to the smoothing step used in G-DICE. For each robot $i$, the weighted log-likelihood function is maximized over $\theta_{q'}^{(i)}$ for KLR training,
\begin{equation}
	l^{(i)}(\theta^{(i)}) = \sum_{b=1}^{N_{KLR}}w_b\log \frac{\exp(\theta_{q'_b}^{(i)T}o^{e(i)}_b)}{\sum_{k=1}^{N_n}\exp(\theta_{q_k}^{(i)T}o^{e(i)}_b)},
\end{equation}
where $o^{e(i)}_b$, ${q'_b}^{(i)}$, and $\theta_{q'_b}^{(i)}$ are transition function training inputs, outputs, and kernel weights for the $b$-th observation bundle. The partial derivative with respect to the $j$-th component of each parameter is,
\begin{equation}
\frac{\partial}{\partial\theta^{(i)}_{q',j}}l(\theta^{(i)})\! =\!\!\!\!\! \sum_{b=1}^{N_{KLR}}\!\!\!\!w_bo^{e(i)}_{b,j}\!\!\left\lbrack\!\mathbb{I}(q'_b\!=\!q')\!-\!\frac{\exp(\theta_{q'}^{(i)T}o^{e(i)}_b)}{\sum_{k=1}^{N_n}\exp(\theta_{q_k}^{(i)T}o^{e(i)}_b)}\!\right\rbrack\!\!,
\end{equation}
where $\mathbb{I(\cdot)}$ is the indicator function. The log-likelihood can be maximized using a quasi-Newton method (our implementation uses the Broyden-Fletcher-Goldfarb-Shanno algorithm). To improve the generalization of the learned model, $L_2$ regularization is used during weight training.

EPSCKO is outlined in \cref{alg:EPSCKO}. It begins by specifying an empty SK-FSA policy and $N_{KLR}$-length FIFO circular kernel basis queue for each robot (\cref{alg:EPSCKO}, Lines~\ref{line:epscko_init_graph}-\ref{line:epscko_init_queue}). The best-value-so-far, $\bar{V}_{b}$, and worst-joint-value, $\bar{V}_{w,0}$, are set to $-\infty$ (\cref{alg:EPSCKO}, Line~\ref{line:epscko_init_V}). To encourage policy space exploration, SK-FSA parameter vectors are initialized such that associated distributions are uniform (\cref{alg:EPSCKO}, Line~\ref{line:init_params_1}). 

The main algorithm loop updates the SK-FSA policy over $N_k$ iterations, using the maximal entropy injection scheme detailed in \cref{subsec:max_entropy_injection} to accelerate search. Entropy injection is initially disabled and a flag indicating successful entropy injection in the current iteration is set to \texttt{False} (\cref{alg:EPSCKO}, Line~\ref{line:allow_ent_inject}). The team's SK-FSA policies are evaluated $N_s$ times, with perceived continuous observation and node transition trajectories saved for KLR training (\cref{alg:EPSCKO}, Line~\ref{line:decposmdp_eval}). MA selections and node transitions from policies exceeding the previous iteration's worst joint value are tracked in $\overline{KLR}_{list}$ (\cref{alg:EPSCKO}, Lines~\ref{line:epscko_reject_bad_policies_1}-\ref{line:epscko_reject_bad_policies_3}). The best-value-so-far, $\bar{V}_b$, is saved (\cref{alg:EPSCKO}, Line~\ref{line:epscko_save_best_policy}). Trajectory lists are pruned to retain only the best $N_b$ trajectories (\cref{alg:EPSCKO}, Line~\ref{line:epscko_retain_best_nb}). Continuous observations and node transitions from this list are pushed to the FIFO queue, causing old trajectories to be popped (\cref{alg:EPSCKO}, Line~\ref{line:epscko_append_oe_hist}). The iteration's worst joint value, $\bar{V}_{w,k+1}$, is then updated.

At this point, the algorithm checks if the Dec-POSMDP joint value has converged. If so, entropy injection is enabled to counter convergence to a local optima (\cref{alg:EPSCKO}, Line~\ref{line:epscko_allow_injection}). This does not imply entropy injection \emph{will} occur, only that it is \emph{allowed to} occur. Each robot subsequently updates its MA distribution parameter vector, $\theta^{(i)(\pi|q)}$, using a smoothed MLE approach (\cref{alg:EPSCKO}, Lines~\ref{line:epscko_mle_update}-\ref{line:epscko_param_smooth}). As discussed earlier, weighted log-likelihood maximization is used to train the KLR model for each node transition function (\cref{alg:EPSCKO}, Line~\ref{line:epscko_klr_update}). 

Next, if maximal entropy injection is allowed, entropies of sampling distributions are calculated and (if necessary) injection occurs (\cref{alg:EPSCKO}, Line~\ref{line:epscko_try_inject}). As transition function $\delta^{(i)}$ is continuous and non-linear, an approximate measure of its entropy is calculated using transition distributions sampled at its underlying set of observation kernels. This approximation was found to work well in practice (\cref{subsec:expts_epscko}) and is computationally efficient as it avoids domain re-sampling. To increase entropy of the node transition function, a continuous uniform distribution injection is done using update rule \cref{eq:entropy_injection}. 
If entropy injection is conducted for any robot, the current iteration's worst joint value, $\bar{V}_{w,k+1}$, is set to $-\infty$ (\cref{alg:EPSCKO}, Line.~\ref{line:epscko_reset_worst}). This critical step ensures trajectories sampled in the next iteration can actually be used for policy exploration.

EPSKCO is an anytime algorithm applicable to continuous-observation Dec-POMDPs and Dec-POSMDPs. This approach also offers memory advantages to discretization as SK-FSA memory usage is $O(N_{KLR}N_bN_n|\mathbb{\bar{\breve{O}}}|)$, in contrast to $O(d^{|\mathbb{\bar{\breve{O}}}|}N_n^2)$ for FSAs with discretization resolution $d$.

\section{Experiments} \label{sec:expts}
This section first validates maximal entropy search acceleration, which resolves a long-standing convergence issue for sampling-based Dec-POSMDP algorithms. Then, EPSCKO is evaluated against discrete approaches in the first ever continuous-observation Dec-POMDP/Dec-POSMDP domain. 

\subsection{Accelerated Policy Search}\label{subsec:expts_accel}
\begin{figure}[t]
	\centering
	\includegraphics[trim={0.2cm 2.7cm 0.9cm 0.3cm},width=1\linewidth]{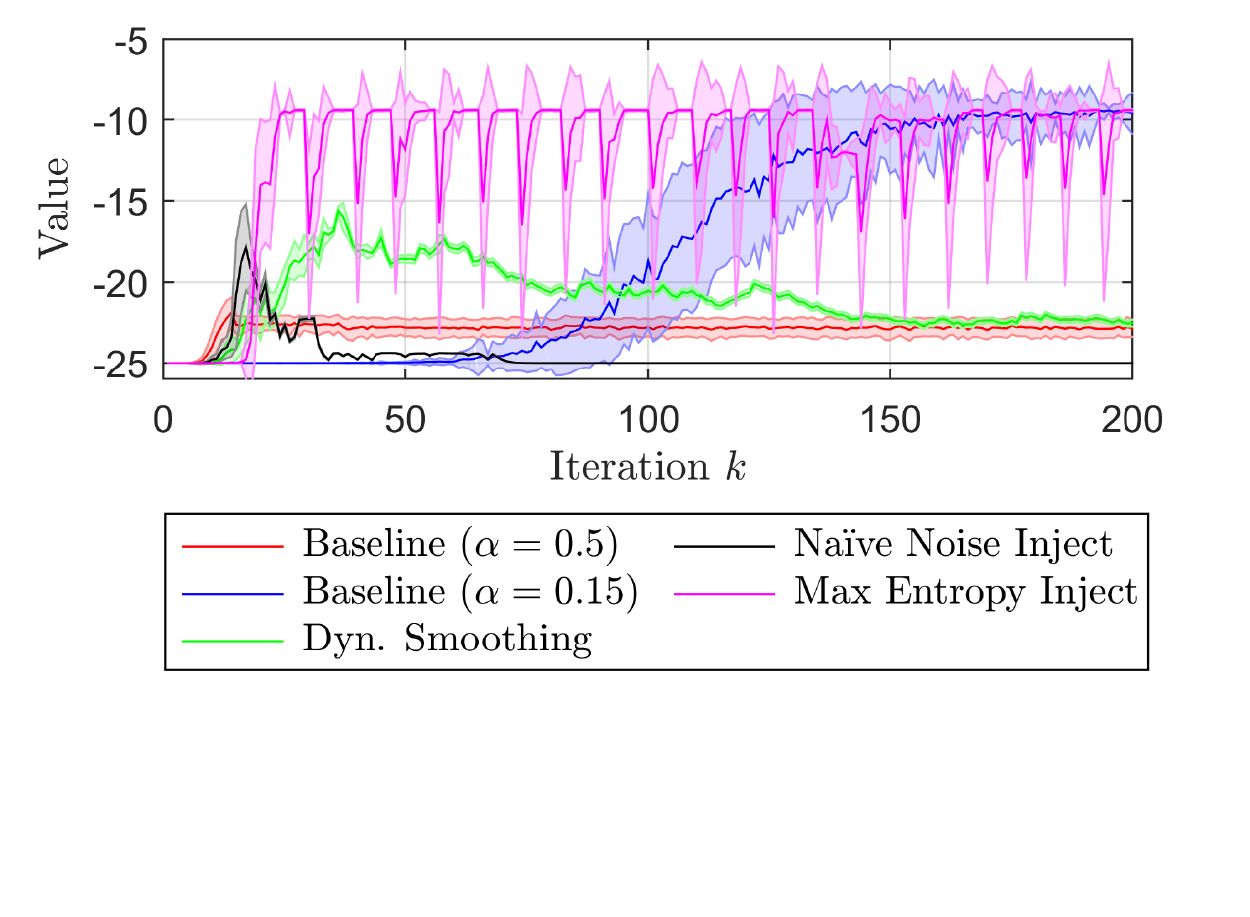}
	\caption{Comparison of search acceleration approaches for NAMO domain, using a $N_n=5$ node policy.}\label{fig:gdice_variants_compare}
\end{figure}

We evaluate policy search acceleration approaches discussed in \cref{sec:max_ent_injection} on the benchmark Navigation Among Movable Obstacles (NAMO) domain \cite{stilman2005navigation} with horizon $h=25$ and a $6 \times 6$ grid. \cref{fig:gdice_variants_compare} shows convergence trends for all approaches. A low learning rate of $0.15$ is needed in G-DICE \cite{Omidshafiei16_ICRA} to find the optimal policy (taking $k=200 $ iterations). 50 policies are sampled per iteration, with 1000 trajectories used to approximate policy value in each iteration, so $1e7$ total policy evaluations are conducted. This computationally expensive evaluation becomes prohibitively large as domain complexity  grows. Increasing learning rate to $\alpha= 0.5$ causes fast convergence to a sub-optimal solution, after which exploration stops due to sampling distribution degeneration. 

Existing search acceleration approaches are also evaluated. Dynamic smoothing with a moderate baseline rate ($\alpha_0 = 0.5$, $\beta = 15$) slightly improves value. However, decay rate $\beta$ is static with no closed-loop feedback from underlying sampling distributions. The result is a sub-optimal policy (found around iteration $k = 35$) which then quickly converges to the same value as the baseline approach with $\alpha=0.5$. Linearly decreasing noise injection with $\omega_{\rm max} = 0.02$ and $r=2000^{-1}$ performs similarly, with fast initial increase in value and subsequent degeneration to a sub-optimal policy. 

The proposed entropy injection method significantly outperforms the above approaches. The same baseline learning rate as previous methods ($\alpha=0.5$) is used with a 3\% entropy injection rate, resulting in much faster convergence (around $k=20$). Sensitivity to $\alpha$ and injection rate is low as value convergence monitoring is conducted in all iterations. While some initial tuning of entropy injection rate is necessary, the key insight is that post-tuning results converge much faster and are more conducive to additional experimentation and analysis (e.g., with domain/policy structure). Oscillations in plots are due to post-convergence injections, which reset underlying sampling distributions and forces further policy space exploration. In practice, the best policy found in a fixed number of iterations would be returned by the algorithm.

\subsection{Continuous Observation Domain}\label{subsec:expts_epscko}
\begin{figure*}[t!]
	\centering
	\begin{minipage}[t]{.2\linewidth}
		\centering
		\includegraphics[trim={2.3cm 10cm 10cm 5cm}, width=1\linewidth]{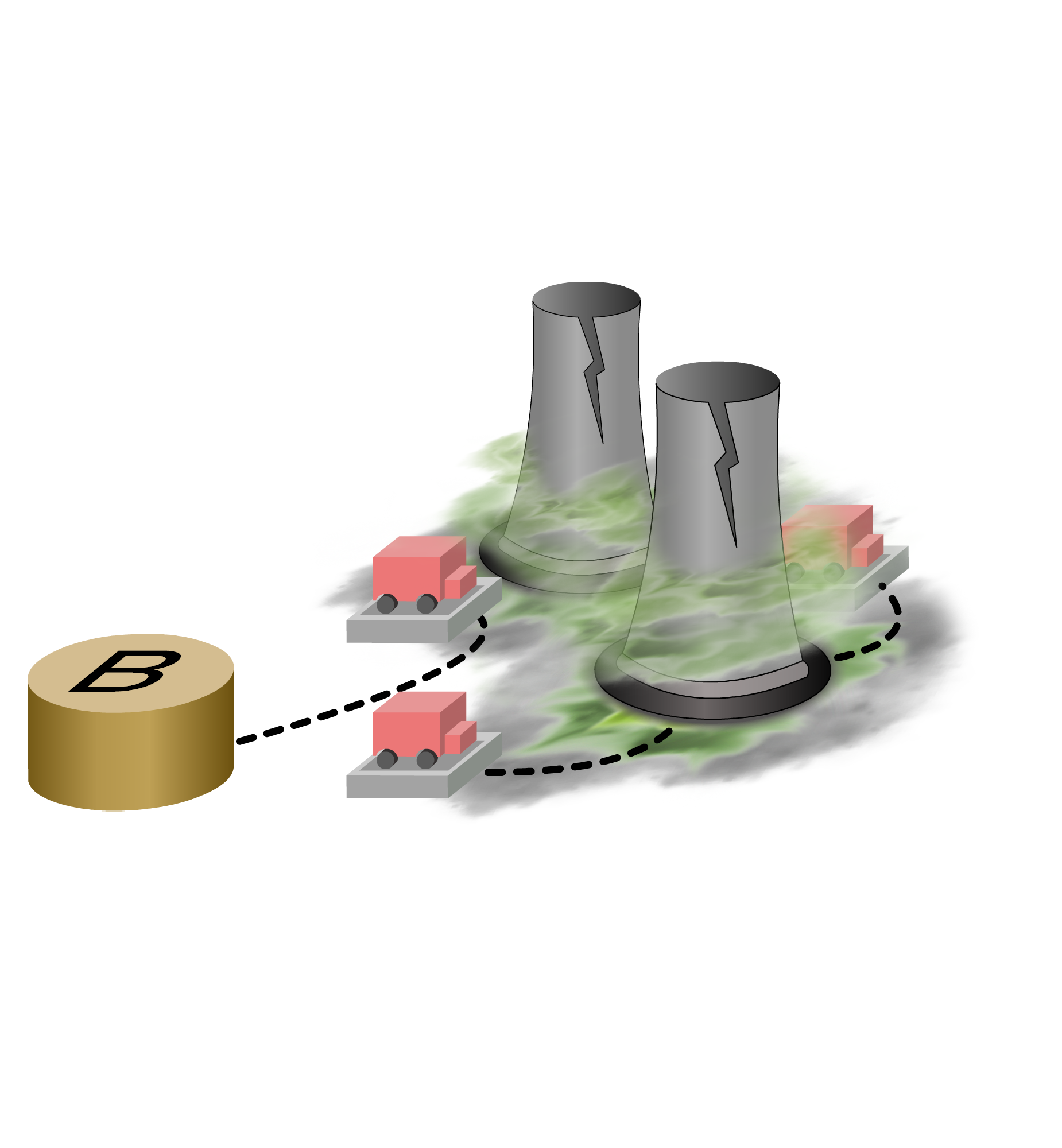}
		\vspace{-12pt}				
		\subcaption{Domain overview (artist's conception).}
		\label{fig:domain_overview_3d_final}
	\end{minipage}%
	\hspace*{\fill}
	\begin{minipage}[t]{.78\linewidth}
		\centering
		\includegraphics[clip,trim={2.7cm 2.35cm 2.3cm 0.53cm}, width=1\linewidth]{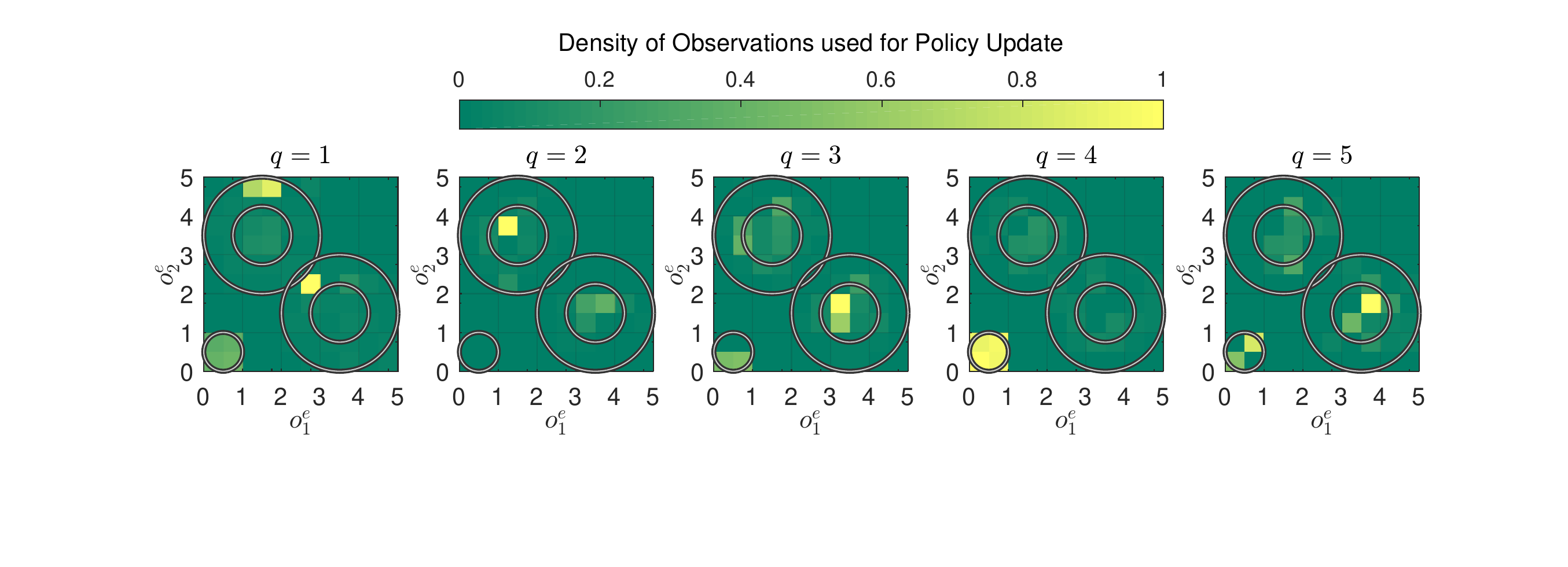}
		\vspace{-12pt}				
		\subcaption{Density of observations used to update discrete Dec-POSMDP policy. $N_n=5$ node policy case with discretization factor $d=10$. Numerous low density bins present due to fine discretization.}\label{fig:sampled_obs_plot_nbins_10}
	\end{minipage}
	\\
	\vspace{-1pt}
	\centering
	\begin{minipage}[t]{.2\linewidth}
		\centering
		\includegraphics[clip, trim={0 28cm 0 33cm}, width=0.925\linewidth]{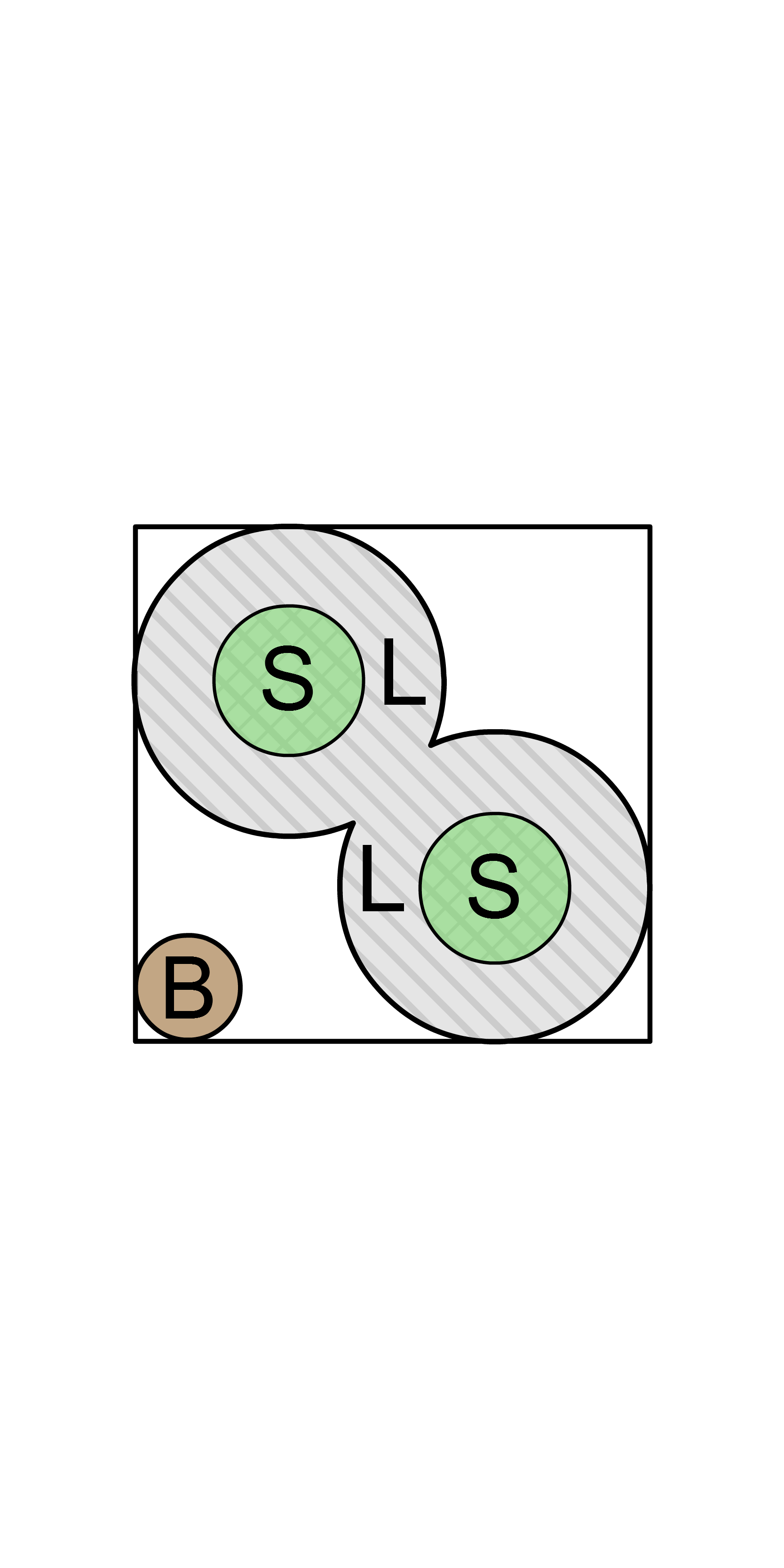}
		\subcaption{Domain overview (key continuous regions).}
		\label{fig:domain_key_regions}
	\end{minipage}%
	\hspace*{\fill}
	\begin{minipage}[t]{.78\linewidth}
		\centering
		\includegraphics[clip, trim={2.5cm 2.35cm 2.5cm 2.55cm}, width=1\linewidth]{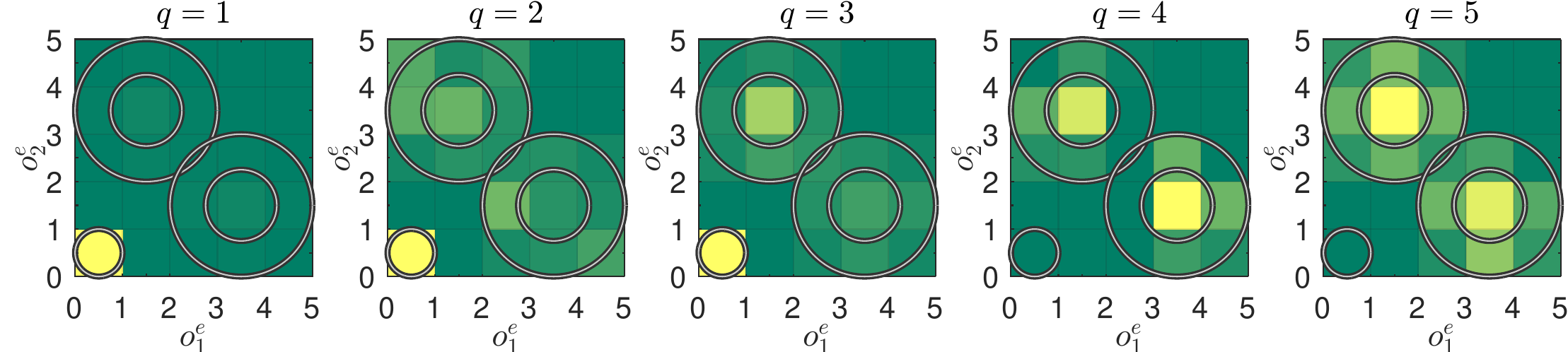}
		\vspace{-12pt}
		\subcaption{Density of observations used to update a discrete Dec-POSMDP policy. $N_n=5$ node policy case with discretization factor $d=5$. Observation density increases in the low discretization resolution case.}\label{fig:sampled_obs_plot_nbins_5}
	\end{minipage}
	\vspace{-5pt}
	\caption{Continuous-observation nuclear contamination domain overview and corresponding discrete policy results.}
\end{figure*}

To evaluate EPSCKO, a multi-robot continuous-observation nuclear contamination domain is considered (\cref{fig:domain_overview_3d_final}). This first-ever continuous-observation Dec-POMDP/POSMDP domain involves 3 robots cleaning up nuclear waste. MAs are \emph{Navigate to base}, \emph{Navigate to waste zone}, \emph{Correct position}, and \emph{Collect nuclear contaminant}. Following MA execution, each robot receives a noisy high-level observation $o^e$ of its 2D ($x,y$) state. The above MAs have non-deterministic durations and a 30\% failure probability (due to nuclear contaminant degrading the robots). This causes poor performance of observation-agnostic policies which memorize chains of MAs, rather than make informed decisions using the observations. 

Robots are initially at the base and must first navigate to the waste zone prior to collection attempt. Robots which execute the \emph{Navigate to base} MA terminate with a random continuous state in a region centered on the base (brown region marked `B' in \cref{fig:domain_key_regions}). The \emph{Navigate to waste zone} MA results in a random terminal state within two large regions surrounding the nuclear zone (everything \emph{interior} of gray regions marked `L' in \cref{fig:domain_key_regions}, \emph{including} the green regions marked `S'). Collection attempts are only possible if the robot is \emph{within} the waste zone (green regions marked `S' in \cref{fig:domain_key_regions}). Collections attempted outside these small contamination regions result in wasted time, which further discounts the team's future joint rewards. Robot can attempt a \emph{Correct position} MA, which re-samples their state to be within these smaller regions. However, repeated attempts may be necessary due to the 30\% MA failure probabilities. 

After successful collection, each robot must return to the base to deposit the waste before attempting another collection. Each collection results in $+1$ joint team reward (with discount factor $\gamma = 0.9$). This domain is particularly challenging due to the high failure rate of MAs, and the presence of a continuous, non-linear decision boundary in the nuclear zone center, where the trade-off between the correction and collection MAs must be considered by robots given their noisy observations. 

\begin{figure}[t]
	\centering
	\includegraphics[trim={0.5cm 0.2cm 1cm 0.3cm},width=1\linewidth]{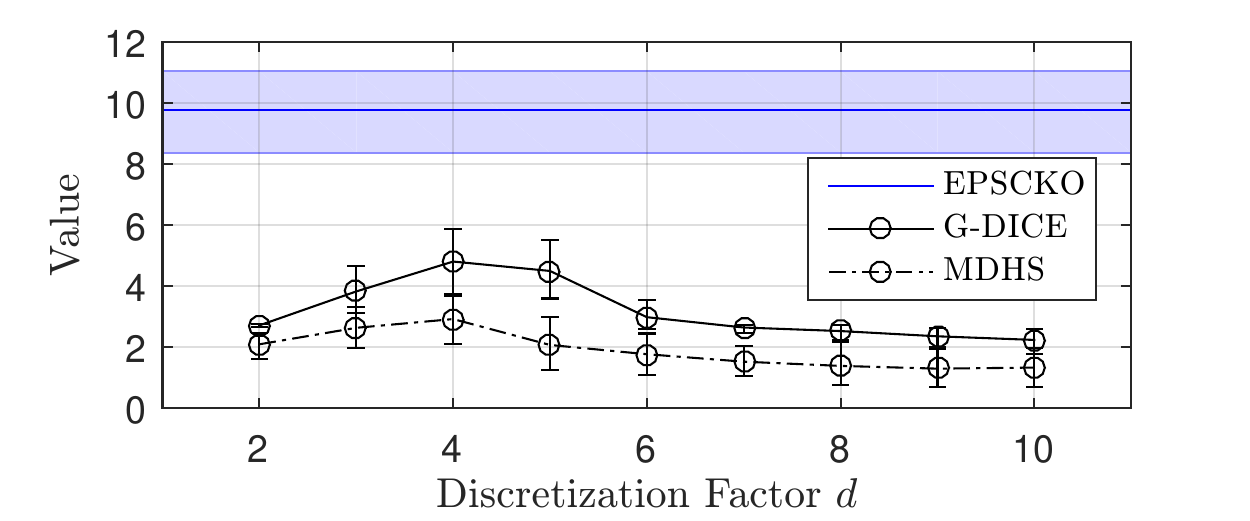}
	\caption{Comparison of discrete and continuous policy search approaches for nuclear contamination domain.}	\label{fig:discrete_continuous_policy_value_comparisons}
\end{figure}

\cref{fig:discrete_continuous_policy_value_comparisons} compares best values obtained using continuous-observation and discrete-observation policy search (EPSCKO, G-DICE with maximal entropy injection, and MDHS). Time horizon $h=40$ was used for evaluation, with each MA taking an average of $1$-$4$ time units to complete. $N_n=6$ nodes were used for both discrete and continuous policies. G-DICE and MDHS results are shown for observation discretization factors $d \in \{2,\ldots,10\}$, with uniform discretization in each observation dimension. EPSCKO significantly outperforms the discrete approaches, more than doubling the mean policy value of the best discrete-observation case ($d=4$). MDHS faces the policy expansion issues discussed in \cref{subsec:decposmdp_solns}. 

G-DICE policy values initially increase with higher discretization resolutions ($d=2$ to $d=4$), yet a drop-off occurs beyond $d=5$. While initially counterintuitive, as higher discretization factors imply increased precision regarding important decision boundaries in the continuous domain, \cref{fig:sampled_obs_plot_nbins_10,fig:sampled_obs_plot_nbins_5} reveal the underlying problem. These plots show the normalized count of observation samples used to compute $N_n=5$ node discrete policies for the $d=10$ and $d=5$ cases, with discounting of old observation samples using \eqref{eq:gdice_smoothing}. In other words, they provide a measure of discrete observation bins which have informed each G-DICE policy throughout its iterations. The core issue for discrete policies is that no correlation exists between decisions at nearby observation bins. Fine discretization meshes, as in \cref{fig:sampled_obs_plot_nbins_10}, result in cyclic processes where observation bins with no previous samples are encountered, therefore causing the robot to make a poor MA selection. Nearby observation bins do not inform the robot during this process, leading it to repeatedly make incorrect decisions. This issue is especially compounded in this domain due to delays caused by high MA failure probabilities, which reduce the overall number of observations received by robots. The result is a highly uninformative policy with no observations made in many bins, in contrast to policies with lower discretization factor (\cref{fig:sampled_obs_plot_nbins_5}).

\begin{figure*}[t]
%
	\vspace{-6pt}
	\centering
	\includegraphics[clip,trim={4.8cm 1.1cm 3.5cm 1.55cm},width=1\linewidth]{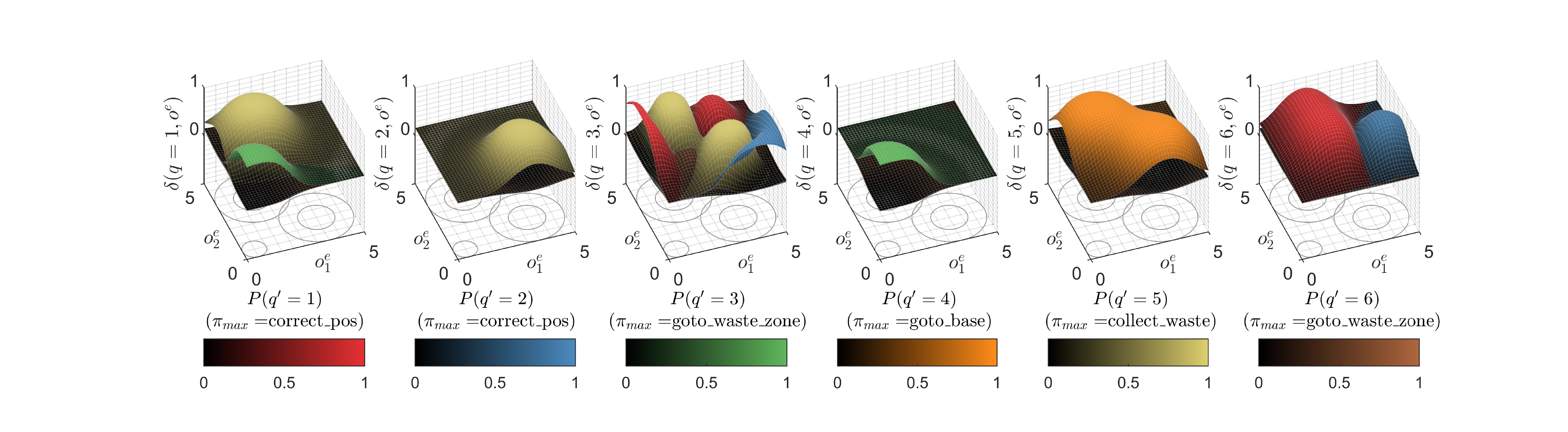}
	\vspace{-12pt}
	\caption{Visualization of an $N_n=6$ node SK-FSA policy transition functions for nuclear contamination domain. For each node $q$ and observation $o^e$, colored 3D manifolds represent probabilities of transitioning to next-nodes $q'$. Colorbars indicate the color associated with each node, as well \emph{highest-probability} MA, $\pi_{\rm max}$, executed in it.}\label{fig:continuous_policy_transition_probas_v2}
	\vspace{-15pt}
\end{figure*}

%
%

To build intuition on continuous-policy decision-making, \cref{fig:continuous_policy_transition_probas_v2} plots transition functions for a 6-node EPSCKO policy. For each node $q$, colored 3D manifolds represent probabilities of transitioning to next-nodes, $q'$, given a continuous observation. Circles plotted beneath transition functions indicate base and nuclear zone locations. Colorbars indicate the transition manifold color associated with each node and the \emph{highest-probability} MA, $\pi_{\rm max}$, executed in it.

Consider a robot policy starting at node $q=1$ (far left in \cref{fig:continuous_policy_transition_probas_v2}) which has two major manifolds (beige and green). Observations under a prominent green manifold region indicate high probability of transitioning to node $q'=3$ (as its colorbar is green), which has $\pi_{\rm max} = $ \emph{Navigate to waste zone}. For $q=1$, this green manifold is centered on the base, which makes intuitive sense as the \emph{Navigate to waste zone} MA should only be executed if the robot is confident it is at the base. Thus, the robot most likely transitions to node $q=3$, and a complex transition function manifold is encountered. Two beige peaks are centered on the small inner regions of the nuclear zone, indicating transition to node $q'=5$, which has $\pi_{\rm max}=$ \emph{Collect nuclear contaminant}. Thus, when the robot is in $q=3$ and confident that it is in the center of the nuclear zone, it attempts a collection MA. Yet, for observations outside the inner nuclear zone, the red and blue manifolds are most prominent. These indicate high probabilities of transitioning to $q'=1$ and $q'=2$, which have $\pi_{\rm max}=$ \emph{Correct position}. Thus, the robot most likely performs a heading correction before continuing policy execution and attempting waste collection. This process continues indefinitely or until the time horizon is reached. Recall that SK-FSA policies are stochastic, so these discussions provide an intuition of the `most likely' continuous-policy behaviors.


\section{Conclusion} \label{sec:conclusion}
This paper presented an approach for solving continuous-observation multi-robot planning under uncertainty problems. Entropy injection for policy search acceleration was presented, targeting convergence issues of existing algorithms, which are exacerbated in the continuous case. Stochastic Kernel-based Finite State Automata (SK-FSAs) were introduced for policy representation in continuous domains, with the Entropy-based Policy Search using Continuous Kernel Observations (EPSCKO) algorithm for continuous policy search. EPSCKO was shown to significantly outperform discrete search approaches for a complex multi-robot continuous-observation nuclear contamination mission---the first ever Dec-POMDP/Dec-POSMDP domain. Future work includes extending the framework to continuous-time planning.



\bibliographystyle{IEEEtran}
\bibliography{TemplateFiles/shayegan_bib,BIB_all/ACL_all,BIB_all/ACL_bef2000,BIB_all/ACL_Publications,references}

\begin{thebibliography}{10}
\providecommand{\url}[1]{#1}
\csname url@samestyle\endcsname
\providecommand{\newblock}{\relax}
\providecommand{\bibinfo}[2]{#2}
\providecommand{\BIBentrySTDinterwordspacing}{\spaceskip=0pt\relax}
\providecommand{\BIBentryALTinterwordstretchfactor}{4}
\providecommand{\BIBentryALTinterwordspacing}{\spaceskip=\fontdimen2\font plus
\BIBentryALTinterwordstretchfactor\fontdimen3\font minus
  \fontdimen4\font\relax}
\providecommand{\BIBforeignlanguage}[2]{{%
\expandafter\ifx\csname l@#1\endcsname\relax
\typeout{** WARNING: IEEEtran.bst: No hyphenation pattern has been}%
\typeout{** loaded for the language `#1'. Using the pattern for}%
\typeout{** the default language instead.}%
\else
\language=\csname l@#1\endcsname
\fi
#2}}
\providecommand{\BIBdecl}{\relax}
\BIBdecl

\bibitem{Bernstein02}
D.~S. Bernstein, R.~Givan, N.~Immerman, and S.~Zilberstein, ``The complexity of
  decentralized control of {M}arkov decision processes,'' \emph{Math. of Oper.
  Research}, vol.~27, no.~4, pp. 819--840, 2002.

\bibitem{AmatoRSS15_v2}
C.~Amato, G.~Konidaris, A.~Anders, G.~Cruz, J.~How, and L.~Kaelbling, ``Policy
  search for multi-robot coordination under uncertainty,'' in \emph{Robotics:
  Science and Systems XI (RSS)}, 2015.

\bibitem{Omidshafiei15_ICRA}
S.~Omidshafiei, A.-A. Agha-Mohammadi, C.~Amato, and J.~P. How, ``Decentralized
  control of partially observable markov decision processes using belief space
  macro-actions,'' in \emph{Robotics and Automation (ICRA), 2015 IEEE
  International Conference on}.\hskip 1em plus 0.5em minus 0.4em\relax IEEE,
  2015, pp. 5962--5969.

\bibitem{DecPOMDPBook16}
F.~A. Oliehoek and C.~Amato, \emph{A Concise Introduction to Decentralized
  {POMDPs}}.\hskip 1em plus 0.5em minus 0.4em\relax Springer, 2016.

\bibitem{hoey2005solving}
J.~Hoey and P.~Poupart, ``Solving {POMDPs} with continuous or large discrete
  observation spaces,'' in \emph{IJCAI}, 2005, pp. 1332--1338.

\bibitem{porta2006point}
J.~M. Porta, N.~Vlassis, M.~T. Spaan, and P.~Poupart, ``Point-based value
  iteration for continuous {POMDPs},'' \emph{Journal of Machine Learning
  Research}, vol.~7, no. Nov, pp. 2329--2367, 2006.

\bibitem{bai2014integrated}
H.~Bai, D.~Hsu, and W.~S. Lee, ``Integrated perception and planning in the
  continuous space: A {POMDP} approach,'' \emph{The International Journal of
  Robotics Research}, vol.~33, no.~9, pp. 1288--1302, 2014.

\bibitem{brechtel2013solving}
S.~Brechtel, T.~Gindele, and R.~Dillmann, ``Solving continuous {POMDPs}: Value
  iteration with incremental learning of an efficient space representation.''
  in \emph{ICML (3)}, 2013, pp. 370--378.

\bibitem{Omidshafiei16_ICRA}
S.~Omidshafiei, A.-A. Agha-Mohammadi, C.~Amato, S.-Y. Liu, J.~P. How, and
  J.~Vian, ``Graph-based cross entropy method for solving multi-robot
  decentralized pomdps,'' in \emph{Robotics and Automation (ICRA), 2016 IEEE
  International Conference on}.\hskip 1em plus 0.5em minus 0.4em\relax IEEE,
  2016, pp. 5395--5402.

\bibitem{journals/orl/CostaJK07}
A.~Costa, O.~D. Jones, and D.~P. Kroese, ``Convergence properties of the
  cross-entropy method for discrete optimization.'' \emph{Oper. Res. Lett.},
  vol.~35, no.~5, pp. 573--580, 2007.

\bibitem{conf/wsc/BotevK04}
Z.~I. Botev and D.~P. Kroese, ``Global likelihood optimization via the
  cross-entropy method, with an application to mixture models.'' in
  \emph{Winter Simulation Conference}.\hskip 1em plus 0.5em minus 0.4em\relax
  WSC, 2004, pp. 529--535.

\bibitem{Kroese2006}
D.~P. Kroese, S.~Porotsky, and R.~Y. Rubinstein, ``The cross-entropy method for
  continuous multi-extremal optimization,'' \emph{Methodology and Computing in
  Applied Probability}, vol.~8, no.~3, pp. 383--407, 2006.

\bibitem{journals/icga/ThieryS09a}
C.~Thiery and B.~Scherrer, ``Improvements on learning tetris with cross
  entropy.'' \emph{ICGA Journal}, vol.~32, no.~1, pp. 23--33, 2009.

\bibitem{devroye2013probabilistic}
L.~Devroye, L.~Gy{\"o}rfi, and G.~Lugosi, \emph{A probabilistic theory of
  pattern recognition}.\hskip 1em plus 0.5em minus 0.4em\relax Springer Science
  \& Business Media, 2013, vol.~31.

\bibitem{shannon2001mathematical}
C.~E. Shannon, ``A mathematical theory of communication,'' \emph{ACM SIGMOBILE
  Mob. Comp. and Comm. Rev.}, vol.~5, no.~1, 2001.

\bibitem{amato2010optimizing}
C.~Amato, D.~S. Bernstein, and S.~Zilberstein, ``Optimizing fixed-size
  stochastic controllers for {POMDPs} and decentralized {POMDPs},''
  \emph{Auton. Agents and Multi-Agent Sys.}, vol.~21, no.~3, pp. 293--320,
  2010.

\bibitem{oliehoek2010value}
F.~Oliehoek, \emph{Value-based planning for teams of agents in stochastic
  partially observable environments}.\hskip 1em plus 0.5em minus 0.4em\relax
  Amsterdam University Press, 2010.

\bibitem{bernstein2009policy}
D.~S. Bernstein, C.~Amato, E.~A. Hansen, and S.~Zilberstein, ``Policy iteration
  for decentralized control of markov decision processes,'' \emph{J. of Artif.
  Intell. Res.}, vol.~34, no.~1, p.~89, 2009.

\bibitem{carden2014convergence}
S.~W. Carden, ``Convergence of a {Q}-learning variant for continuous states and
  actions,'' \emph{J. of Artif. Intell. Res.}, vol.~49, pp. 705--731, 2014.

\bibitem{zhu2012kernel}
J.~Zhu and T.~Hastie, ``Kernel logistic regression and the import vector
  machine,'' \emph{Journal of Computational and Graphical Statistics}, 2012.

\bibitem{stilman2005navigation}
M.~Stilman and J.~J. Kuffner, ``Navigation among movable obstacles: Real-time
  reasoning in complex environments,'' \emph{International Journal of Humanoid
  Robotics}, vol.~2, no.~04, pp. 479--503, 2005.

\end{thebibliography}

\end{document}